\documentclass[conference]{IEEEtran}

\usepackage{cite}      

\usepackage{graphicx}  
\begin{document}


\title{Dual frequency 230/690 GHz interferometry at the Submillimeter Array}

\author{\authorblockN{Todd R. Hunter, John W. Barrett, Raymond Blundell, 
Robert D. Christensen,
Robert S. Kimberk,\\ Steven P. Leiker, Daniel P. Marrone, 
Scott N. Paine, D. Cosmo Papa, Nimesh Patel, Patricia Riddle, \\
Michael J. Smith, T.K.~Sridharan, C.-Y. Edward Tong, Ken H. Young
and Jun-Hui Zhao}
\authorblockA{Harvard-Smithsonian Center for Astrophysics\\
60 Garden Street\\
Cambridge, MA 02138 USA\\
Email: thunter@cfa.harvard.edu}
}


%


\maketitle

\begin{abstract}

The Submillimeter Array (SMA), a collaboration between the Smithsonian
Astrophysical Observatory and the Academica Sinica Institute for
Astronomy and Astrophysics of Taiwan, is an eight-element
radio-interferometer designed to operate throughout the major
atmospheric windows from about 180 to 900~GHz.  In an effort to
mitigate the effects of atmospheric instabilities which limit the
phase coherence of the array especially in the higher frequency bands,
the array was designed to allow simultaneous operation of a low
frequency receiver ($<350$~GHz) with a high frequency receiver
($>330$~GHz).  The overlap region of 330-350~GHz was included to
facilitate dual polarization measurements in the frequency range
considered to offer the highest sensitivity for continuum observations
with the array.
 
So far, the array is equipped with working SIS receivers covering the
frequency ranges 176-256~GHz, 260-350~GHz, and 600-700~GHz, and single
frequency operation has been routine in the lower two frequency bands
for the past year.  More recently, with the completion of IF hardware
required to make full use of the SMA cross-correlator, dual receiver
operation became possible.  We have since made a number of Galactic
and extra-galactic astronomical observations in dual-band mode with
the hopes of using the 230~GHz receiver as a phase reference to enable
improved interferometry in the 650~GHz band.  We will present the
current antenna and receiver performance, some of the first
interferometric images in the 650~GHz receiver band, and our initial
attempts at phase transfer.

\end{abstract}


%
\IEEEpeerreviewmaketitle

\section{Introduction}

The Submillimeter Array (SMA), a collaboration between the Smithsonian
Astrophysical Observatory and the Academica Sinica Institute for
Astronomy and Astrophysics, is an eight-element, heterodyne
radio-interferometer designed to operate throughout the major
atmospheric windows from about 180 to 900~GHz.  It is the first and
only imaging interferometer to operate in the submillimeter band.  The
general design and operation of the SMA has been described previously
\cite{Blundell05}.  Located near the 4200-meter summit of Mauna Kea,
Hawaii, the array was dedicated in November 2003 and initial science
results were collected in a 2004 special issue of {\it The
Astrophysical Journal} \cite{Ho04}.  The first SMA results at 690~GHz
were published in 2004 based on observations of the carbon star
IRC+10216 obtained with only three antennas \cite{Young04}.  Since
that time, the second IF path of the SMA has been built and commissioned
and improvements to the 690~GHz receivers and local oscillator (LO)
control hardware has enabled fully-remote, dual-frequency operation
for the first time.  With this capability, the prospects for ``phase
referenced calibration'' (i.e.  scaling the measured atmospheric phase
from a low frequency band to a high frequency band) can now be
evaluated.  This paper summarizes the present system and presents the
first results obtained during a special observing campaign during
January-February 2005.

 

\section{System Details}

\subsection{Antenna performance}

The SMA 6-meter antennas are composed of 72 panels with 4 adjusting
screws per panel.  The surface is accurately measured (to $8~\mu$m
RMS) via near-field holography and the required screw adjustments to
correct the surface are computed \cite{tks}.  In preparation for
650~GHz work, all antennas were adjusted in this manner to improve the
surface accuracy.  As of February 2005, all antennas were better than
$22~\mu$m RMS with the best antenna being $14~\mu$m RMS (see Table
Table~\ref{rms}).  Confirmation of the antenna surface accuracy,
especially at elevations different from the beacon (i.e. $>20$~deg),
has been obtained via aperture efficiency measurements on celestial
objects (planets).  


\begin{table}
\centering
\caption{Antenna performance as measured by holographic surface RMS (left column), 
and by aperture efficiency scans on celestial objects (right columns). }
\begin{tabular}{|c|c|cc|}
\hline
Antenna & RMS surface & \multicolumn{2}{c|}{Aperture efficiency (\%)} \\ 
Number  &  ($\mu$m)   & 230 (GHz) & 345 GHz\\
\hline 
1 & 15 & 77 & 68 \\
2 & 14 & 74 & 67 \\
3 & 15 & 75 & 61 \\
4 & 18 & 73 & 64 \\
5 & 20 & 75 & 60 \\
6 & 22 & 74 & 60 \\
\hline
\end{tabular}
\label{rms}
\end{table}


\subsection{Optical alignment}

In order to use two receivers simultaneously, the mixer feedhorns must
be aligned on the sky to within a fraction of the primary beamsize at
the highest frequency ($17''$ at 690~GHz for the SMA antennas).
Alignment of the receiver feeds is done using near-field vector beam
measurements in the laboratory \cite{tong}.  This method produces
beams aligned to within $5''$ on the sky, as demonstrated by pointing
calibration performed in single-dish mode on each SMA antenna.
Table~\ref{feedoffsets} lists the feed offset terms measured on each
of the six SMA antennas used during the recent 650~GHz campaign.  The
feed offsets are modelled by two parameters: the radial misalignment
$r$ in arcseconds, and the phase angle $\phi$ of this misalignment in
degrees.  These parameters are combined into two terms: $A_1 =
rcos(\phi)$ and $A_2 = rsin(\phi)$, which translate into azimuth and
elevation offsets: azoff = $A_1$sin(Elev) + $A_2$cos(Elev), eloff =
$A_1$cos(Elev) - $A_2$sin(Elev).  The repeatability of the feed offset
measurements is $1-2''$.  During dual-frequency observations, the
antennas are pointed such that the high frequency receiver is aimed at
the target.  The largest resulting pointing error for the 230~GHz feed
($7''$) corresponds to less than $5\%$ loss due to primary beam 
attenuation.

\begin{table}
\centering
\caption{Feed offset terms measured on the SMA antennas.   }
\begin{tabular}{|c|cc|cc|}
\hline
Antenna & \multicolumn{2}{c|}{345 Rx vs 230 Rx} & \multicolumn{2}{c|}{690 Rx vs 230 Rx} \\
Number & A$_1$ ($''$) & A$_2$ ($''$) & A$_1$ ($''$) & A$_2$ ($''$)\\
\hline 
1 & $-0.2$ & +6.1 & $-2.7$ & +7.0 \\
2 & +4.3 & +2.1 & $-0.85$ & +1.6\\  
3 & $-0.7$ & $-1.1$ & +4.1 & $-0.8$ \\  
4 & $-5.3$ & +4.9 & $-2.2$ & +3.6\\
5 & $-3.9$ & $-1.3$ & +4.8 & $-3.4$\\
6 & $-0.1$ & +4.6 & +4.4 & +3.4\\ 
\hline
\end{tabular}
\label{feedoffsets}
\end{table}

\subsection{Mixer performance}

\begin{figure}
\centering
\includegraphics[width=3.5in]{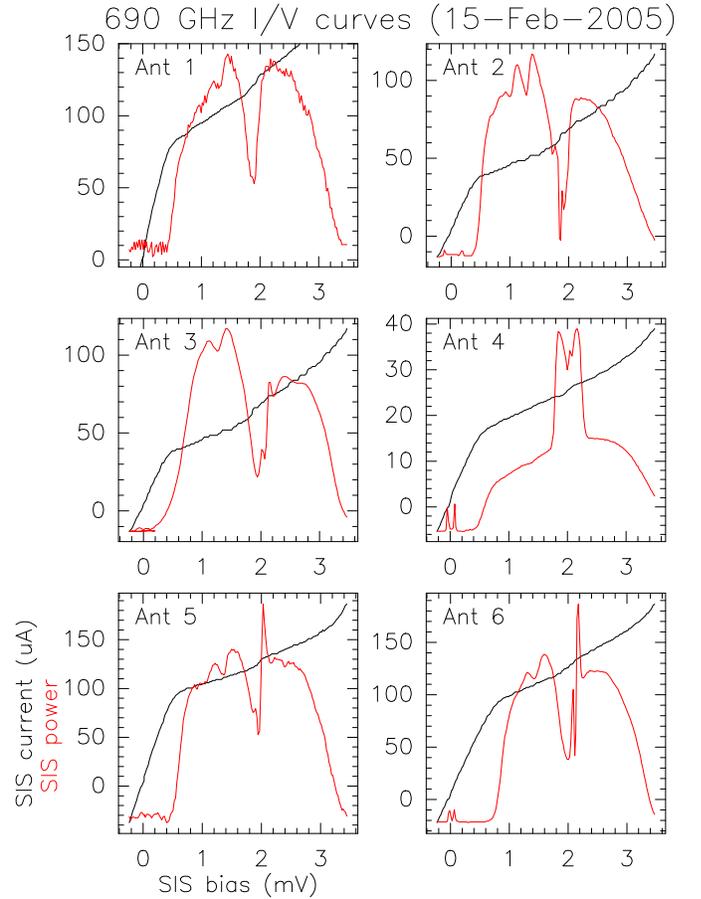} 
\caption{SIS mixer current vs. bias voltage (black line) and SIS mixer 
power vs. bias voltage (red line) characteristics of the
650~GHz receivers on the SMA as measured remotely in the field. 
The SIS power is plotted in arbitrary units.}
\label{ivcurves}
\end{figure}

\begin{figure}
\centering
\includegraphics[width=2.5in,angle=-90]{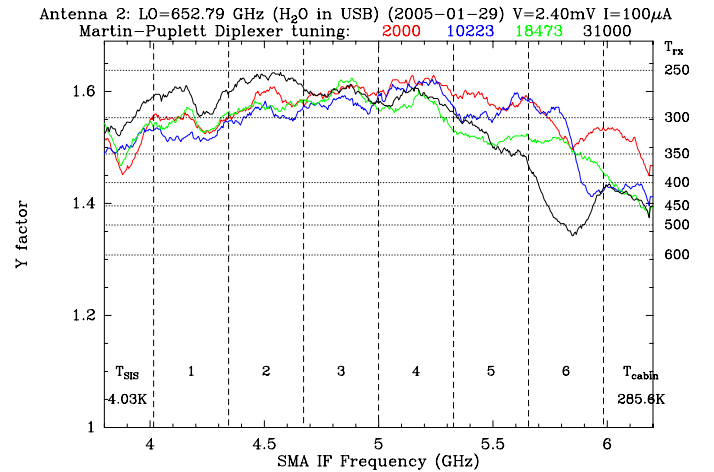} 
\caption{Receiver temperature as a function of the IF passband (x-axis) and 
the Martin-Puplett diplexer tuning positions (colored lines) as measured 
with an HP8563 spectrum analyzer in the field in one of the SMA antennas.  
The data were obtained using the traditional y-factor method by taking the
ratio of two traces: one observing an ambient temperature load and one
observing a liquid nitrogen temperature load.}
\label{yfactor}
\end{figure}

\begin{figure}
\centering
\includegraphics[width=2.6in,angle=-90]{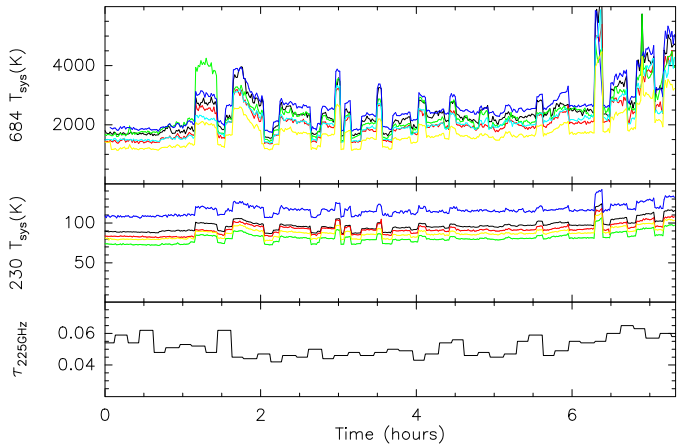}
\caption{System temperatures on each of the six antennas at 684~GHz  (top
panel)
and 230~GHz (middle panel) during the night of Feburary 16, 2005.  The 
maximum elevation
of the target source (Sgr A*) was only 42 degrees above the horizon 
(i.e. 1.5 airmasses).  For reference, the zenith opacity at 225~GHz 
as measured by the tipping radiometer at the Caltech
Submillimeter Observatory is also plotted (lower panel).
}
\label{tsys}
\end{figure}

In all three bands, the SMA receivers contain double-sideband mixers
fabricated with niobium SIS junctions (cooled to 4.2~K) followed by 
second-stage HEMT amplifiers.  The typical I/V characteristics of
these receivers in the field is shown in Figure~\ref{ivcurves}.  While
the instantaneous receiver bandwidth is about 60~GHz, the correlator
can process only 2~GHz in each sideband.  The mixer performance in the
4-6~GHz IF passband was measured in each of the antennas via the
y-factor method.  The passband response was recorded on a spectrum
analyzer, first toward an ambient temperature load, and second toward
a liquid nitrogen temperature load.  A representative plot is shown in
Figure~\ref{yfactor}.  The typical receiver temperature (averaged
across the IF band) was 300-400~K, or about ten times the quantum
limit.  In good dry weather conditions on Mauna Kea ($\tau_{\rm
225GHz} < 0.05$), this level of performance results in system
temperatures as low as 900-1200~K near zenith.  As an example,
Figure~\ref{tsys} shows a plot of the 684~GHz system temperatures and
the 225~GHz zenith opacity vs. time during the nighttime observations
of February 16, 2005.


\subsection{Motorized LOs}

\begin{figure}
\centering
\includegraphics[width=3.5in]{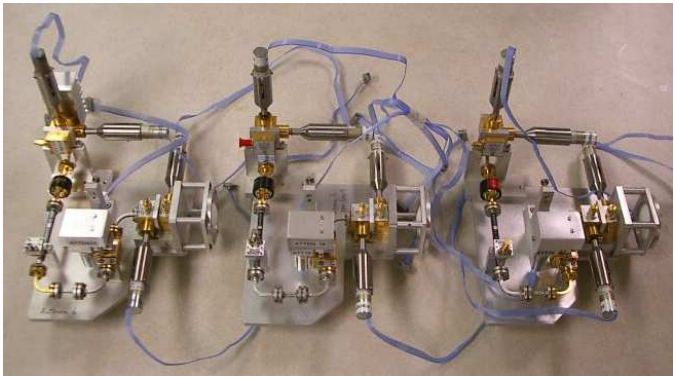} 
\caption{Three of the six motorized 600-700~GHz LO chains installed on
the SMA. At the top is the Gunn oscillator with two linear actuators
at right angles to one another.  Next is a waveguide isolator followed
by a coupler to the PLL harmonic mixer.  The vane attenuator is driven
by a rotary motory via a gear assembly.  The following doubler and tripler
stages each have one linear actuator.  The LO radiation exits through
the lens at the right.  Not shown is the Martin-Puplett diplexer
which injects the LO into the receiver through the cryostat window.}
\label{lochain}
\end{figure}

The automation of the low-frequency (230 and 345~GHz) LO chains has
been described previously \cite{Hunter02}.  The 690~GHz LO chains have
now been automated in a similar fashion.  Each LO chain consists of
six motorized components including a Gunn oscillator (2 motors), a
vane attenuator (1 motor), a doubler (1 motor), a tripler (1 motor)
and Martin-Puplett diplexer (1 motor).  A photo of the LO chain is
shown in Figure~\ref{lochain}.

\subsection{Dual IF operation}

One of the major challenges facing millimeter and submillimeter
interferometers is fluctuations in atmospheric path length caused by
water vapor \cite{lay}.  Such path length changes result in phase
changes which ultimately limit the angular resolution and sensitivity
of the instrument.  The standard technique of phase referencing at the
observed frequency can remove slow drifts in atmospheric and
instrumental phase (on timescales of minutes).  However, as one
observes at higher frequencies, the phase drifts increase in
magnitude, the typical system sensitivities degrade, and the number of
calibrators becomes sparse.  All of these factors tend to make the
standard phase referencing technique untenable.  The SMA telescope
offers the benefit of two independent IF paths (via fiber optics) from
the antennas to the correlator which allows simultaneous observations
at two widely separated frequencies, or (eventually) dual-polarization
observations in the 330-350~GHz range.  Observations at different
frequencies enables the possibility of simultaneous phase referencing
from low frequency where the system sensitivity and stability is the
best to high frequency where the tropospheric phase fluctuations are
the largest.  Recently, this concept has been applied successfully in
astronomical very long baseline interferometry (VLBI) experiments by
using non-simultaneous, but rapidly time-interleaved observations at
15, 43 and 86~GHz \cite{Middelberg05}.  This interleaving concept has
also been considered for use in the Atacama Large Millimeter Array
(ALMA) in the future \cite{Daddario04}.  However, the major drawback
of interleaved observations is that it requires frequent retuning of
the receivers which requires some finite amount of time during which
the instrumental and atmospheric phases may change significantly.  By
design, the unique capability of the SMA of {\bf simultaneous}
observations at two frequencies offers the best chance at proving the
utility of phase referencing in the submillimeter regime.

\section{Observations and Results}

\subsection{Conditions and targets}

\begin{figure}
\centering
\includegraphics[width=3.5in]{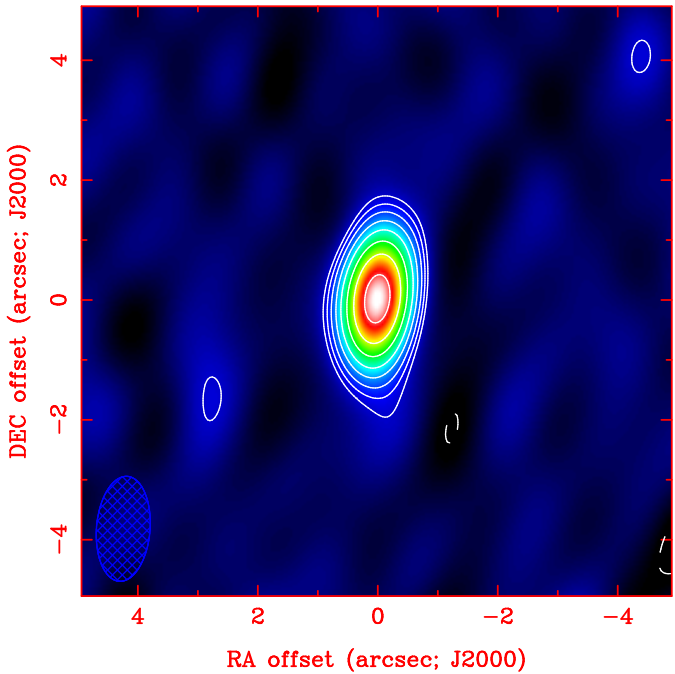}
\caption{Contour plot of the Galactic Center radio source Sgr A* observed
in ~684~GHz continuum emission.  
Levels are -5, -3.5, 3.5, 5, 7, 10, 14, 20, 28, 
and 40$\sigma$, with $\sigma=0.06$~Jy and peak flux density of 2.7~Jy.
One Jy is $10^{-26} {\rm W m}^{-2} {\rm Hz}^{-1}$.
Total integration time is about 3 hours. The synthesized beamwidth of
the interferometer is shown in the lower left corner. The offsets
are relative to the J2000 source position of 17:45:40.0409, $-29$:00:28.118.
} 
\label{sgra} 
\end{figure}

A block of SMA time from January 25 to February 20, 2005 was dedicated
to dual-frequency commissioning and observations.  The first
dual-frequency fringes with five antennas (ten baselines) were
recorded on January 28 on the astronomical maser source W~Hydra.  The
SiO $J$=5-4, $v=1$ maser was observed at 215.596~GHz, and the H$_2$O
$1_{1,0}-1_{0,1}$ maser was observed at 658.007~GHz.  After this
initial success, the sixth antenna became operational and the weather
conditions became favorable for extensive observations of other
astronomical targets from February 14 to 20.  The array was configured
in the ``compact'' formation which provides a synthesized beamwidth of
1.1 by 1.2 arcsecond (for sources of moderate northerly declination).
Unprojected baseline lengths ranged from 16 to 69 meter.  Observations
in the CO $J$=6-5 transition were acquired of the nearby Classical
T~Tauri star TW~Hydra, the protoplanetary nebula CRL618, the high-mass
star-forming region Orion KL, the low-mass star-forming region
IRAS~16293-2422 and the ultraluminous infrared galaxy Arp~220.  The
Galactic Center source Sgr~A* was observed in the continuum at
684~GHz. Finally, the massive star-forming region G240.31+0.07 and the
evolved star VY~Canis~Majoris were observed in the C$^{18}$O $J$=6-5
transition.  In each case, the typical on-source integration times
were three hours.  Detailed results from these observations will be
presented elsewhere.  As an example of the data quality, an image of
the Galactic Center point source Sgr~A* is shown in Figure~\ref{sgra}.
This image was produced via standard phase referencing at 684~GHz
between the target and strong calibrators (in this case, 
Ganymede and Ceres, which were $68^\circ$ and $41^\circ$ distant, 
respectively), followed by phase-only self-cal on the target itself.  

\subsection{Comparison of 230/690~GHz phase}

\begin{figure}
\centering
\includegraphics[angle=-90,width=3.5in]{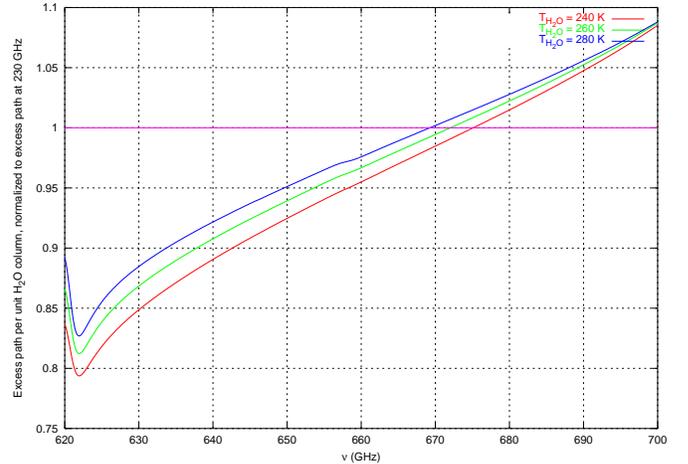}
\caption{Calculation of the theoretical excess path per unit water vapor
column in the 620-700~GHz observing window, normalized to the excess path
at 230~GHz. The three curves correspond to different water vapor 
temperatures in the model. The largest deviation from unity ratio occurs
near water vapor lines, such as the $5_{3,2}-4_{4,1}$ transition at 
620.701~GHz.} 
\label{ratio}
\end{figure}

Ultimately we are interested in whether the phase variations at
$\nu_1=230$~GHz due to fluctuations in the water vapor column above
the antennas can be used to calibrate the proportionally larger phase
variations that these fluctuations cause at $\nu_2=690$~GHz.  To first
order, the expected scaling of the phase variations equals the ratio
of the frequencies (i.e., the non-dispersive case).  This is because
the effect of introducing a small amount of water vapor above one
antenna is to add a (nearly) frequency independent amount of excess
propagation path $L$, which corresponds to a phase delay of
$L/\lambda$ radians where $\lambda = c/\nu$. So the ratio of phases in
the two bands is $(L/\lambda_1)/(L/\lambda_2) = \lambda_2/\lambda_1 =
\nu_1/\nu_2$. In fact, the actual value of $L$ is a function of
frequency, although it only varies significantly near strong
atmospheric water lines \cite{tms,alma404}.  However, because the
650~GHz window sits between two major water lines, the variations are
significant and one should not cancel $L_1/L_2$ in the above formula.
So the actual phase ratio we expect is
$(L_1/L_2)*(\lambda_2/\lambda_1)$.  Figure~\ref{ratio} shows the
theoretical ratio $L_1/L_2$ computed for observations in the 650~GHz
band from the summit of Mauna Kea\footnote{Based on calculations 
performed with the {\it am} atmospheric model \cite{am}, available
at http://cfarx6.cfa.harvard.edu/am}.  The $L_1/L_2$ factor varies 
by about 15-20\%
across the tuning range of the SMA LOs (620-700~GHz), so it is not
negligible.

\begin{figure}
\centering
\includegraphics[angle=0,width=3.5in]{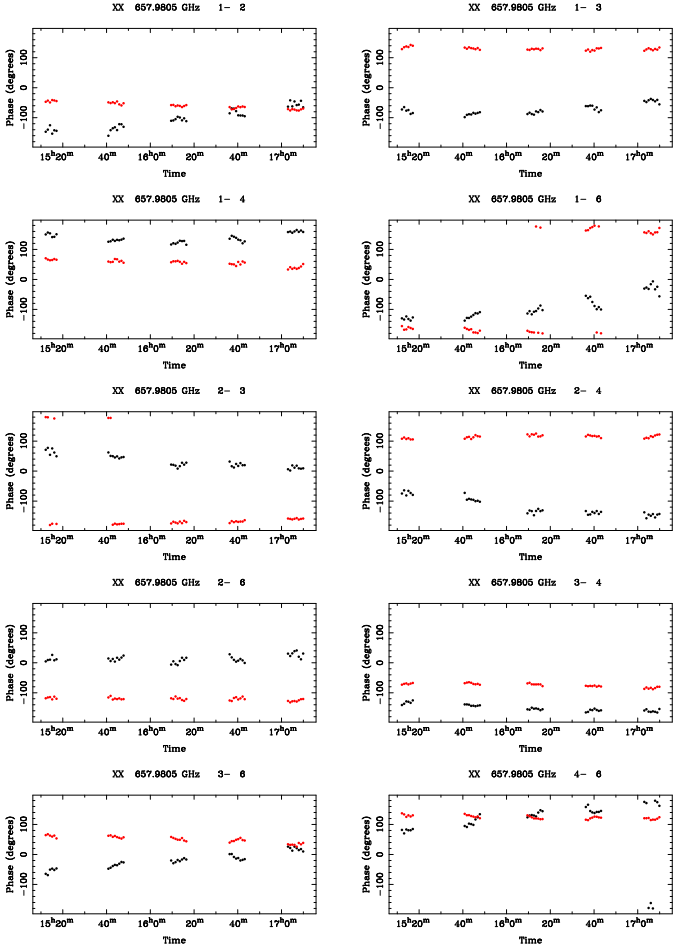} 
\caption{Time series comparison of the spectral line maser phases
of the evolved star W Hydra in the 215~GHz SiO maser (red dots) 
and the 658~GHz H$_2$O maser (black dots).  The ten panels correspond 
to the ten different SMA baselines, as observed with five antennas
on January 28, 2005.
}  
\label{vstime}
\end{figure}

\begin{figure}
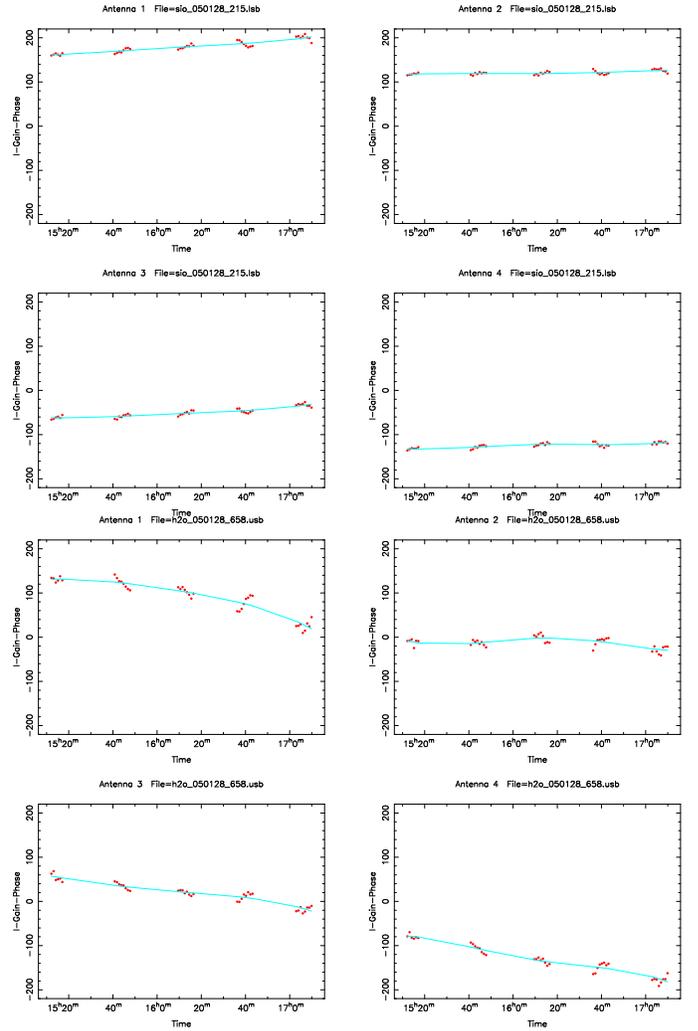

\centering
\includegraphics[angle=-90,width=3.5in]{sioself14.ps}
\includegraphics[angle=-90,width=3.5in]{h2oself14.ps}    
\caption{Point-source antenna-based phase solutions for the SiO 215~GHz 
maser data (upper four panels) and the H$_2$O 658~GHz maser data 
(lower four panels) for 
W Hydra as observed on January 28, 2005. 
The reference antenna was antenna 6 whose phase was defined to be zero
at all times (not shown).  The phase data are plotted against each 
other in Figure~\ref{freqvsfreq}.
} 
\label{selfcal}
\end{figure}

\begin{figure}
\centering
\includegraphics[angle=-90,width=3.5in]{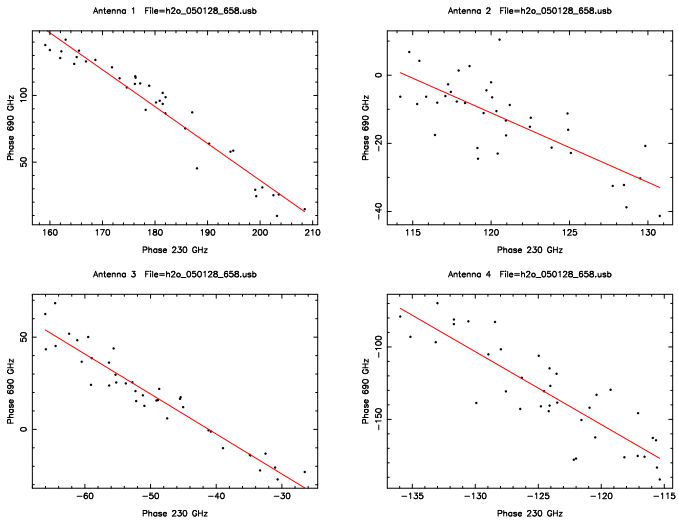} 
\caption{Plot of the 658 GHz phase vs 215 GHz phase data
from Figure~\ref{selfcal}
with linear regression fits overlaid.  The negative slope results from the
fact that the lower sideband SMA data are written with an opposite sign 
convention to the upper sideband SMA data.  On antenna 1, the magnitude 
of the slope is close to the expected theoretical value of -2.90, as seen 
in Table~\ref{linearceres}. 
} 
\label{freqvsfreq}  
\end{figure}

\begin{figure}
\centering
\includegraphics[angle=0,width=3.5in]{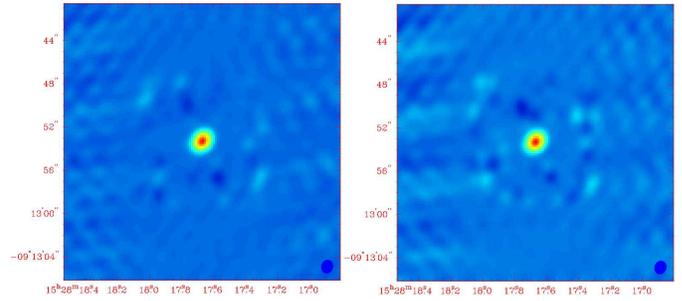}   
\caption{Left image: the image resulting from a standard selfcal solution 
computed from the raw 690 GHz data of Ceres and applied to the same
data. Peak = 10.5~Jy, rms=39~mJy, SNR=267.  Right image: the result 
of taking the standard selfcal solution at 230 GHz, scaling it to 
690~GHz via the slope and offset (Table~\ref{linearceres}), 
and applying it the 690~GHz raw data.  Peak = 9.9~Jy, rms=51~mJy,
SNR=193. The degree to which the images are similar verifies the promise
of this technique.
} 
\label{images}  
\end{figure}

The 230/690 SMA observations recorded on January 28, 2005 have been
analyzed to search for correlations of phase between the two frequency
bands.  Figure~\ref{vstime} shows a time series comparison of the
215~GHz phase averaged across the central six spectral channels of the
SiO maser line with the 658~GHz phase averaged across the central 21
spectral channels of the H$_2$O maser line.  The baseline-based phase
measurements were converted to antenna-based phases using the selfcal
function of Miriad \cite{miriad}.  The gains were derived
independently for each receiver band, using antenna 6 as the reference
antenna, and are shown in Figure~\ref{selfcal}. The gains are plotted
against each other in Figure~\ref{freqvsfreq} using smavarplt function
of Miriad, and the resulting linear regression results are listed in
Table~\ref{linearceres}.  In summary, good correlations are seen on
antennas 1, 2 and 3, with antenna 1 exhibiting a slope closest to the
expected theoretical ratio (based on the model in Figure~\ref{ratio}).
Antenna 4 exhibited anomalous relative phases between the two
frequency bands on many nights of the observing campaign, and clearly
had some instrumental problem.  Nevertheless, the fact that much of
the phase data are well-correlated between the two bands is a
promising sign for future phase referencing at the SMA.  A similar
analysis has been done with the lower-sideband continuum data on 
the minor planet Ceres ($diameter~0.5''$) observed for 3.5 hours 
on February 18, 2005.
The linear regression results are
listed in Table~\ref{linearceres}.  Figure~\ref{images} shows a comparison 
of images of Ceres which demonstrates that phase transfer can be done
successfully between the two receiver bands.  The image in the left panel 
results from running selfcal on the 690~GHz data, and the signal-to-noise 
ratio is 267.  The image in the right panel results from running selfcal
on the 230~GHz data, scaling the solutions to 690~GHz, and applying the 
scaled solutions to the 690~GHz raw data.  In this case, the 
signal-to-noise ratio is 193.


\begin{figure}
\centering
\includegraphics[angle=0,width=3.5in]{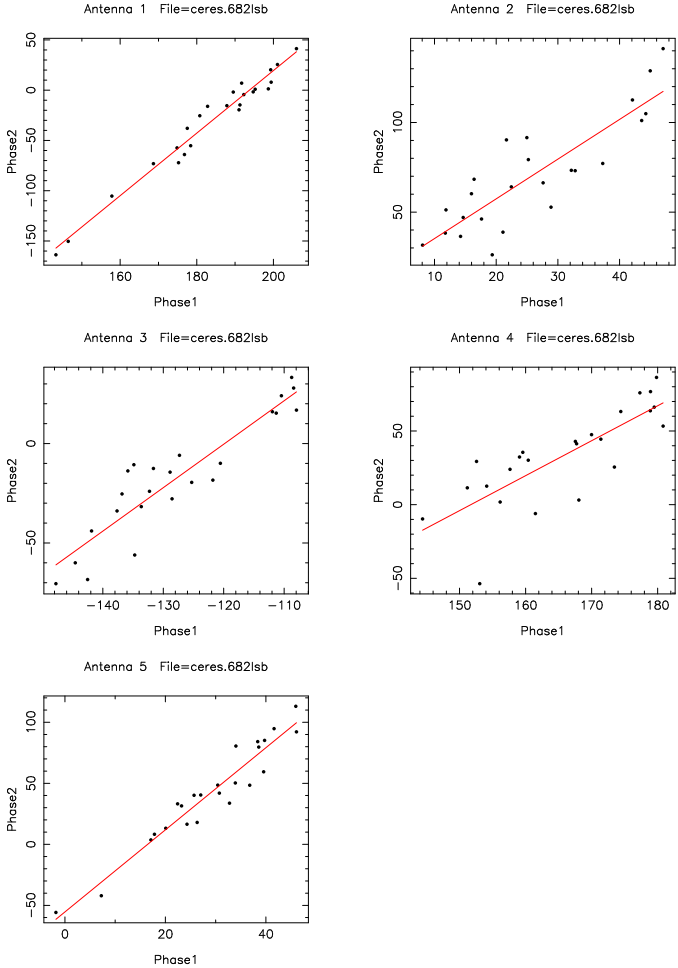} 
\caption{Plot of the 682~GHz vs 221~GHz antenna-based continuum phase 
solutions of Ceres
with linear regression fits overlaid.  On antennas 1 and 5, the magnitude 
of the slope is close to the expected theoretical value of 3.15, as seen 
in Table~\ref{linearceres}.  
} 
\label{ceresfreqvsfreq}  
\end{figure}

\begin{table}
\centering
\caption{Linear regression analysis of the 690~GHz band
phase vs the 230~GHz band phase observed on W Hydra 
and Ceres}
\begin{tabular}{|c|c|c|c|c|}
\hline
Source & Antenna & Sideband & Correlation & Slope \\ 
\hline
& 1 & & 0.97 & $-2.77$ \\
& 2 & & 0.76 & $-2.04$ \\  
W Hydra & 3 & USB & 0.96 & $-2.17$ \\  
(28Jan2005) & 4 & & 0.86 & $-5.03$ \\
& theory &  & 1.00 & $-2.90$ \\
\hline 
& 1 &  & 0.97      & $2.81$ \\
& 2 &  & 0.88      & $2.13$ \\  
Ceres & 3 & USB & 0.95      & $2.08$ \\  
(18Feb2005)& 4 &  & 0.90      & $2.22$ \\
& 5 &  & 0.97      & 3.14 \\
& theory &  & 1.00      & $3.15$ \\
\hline 
& 1 &  & 0.98      & $3.11$ \\
& 2 &  & 0.86      & $2.22$ \\  
Ceres & 3 & LSB & 0.92      & $2.19$ \\  
(18Feb2005)& 4 &  & 0.79      & $2.37$ \\
& 5 &  & 0.96      & 3.36 \\
& theory &  & 1.00 & 3.16 \\
\hline
\end{tabular}
\label{linearceres}
\end{table}

\section{Conclusions}

The SMA has successfully observed celestial sources in two frequency
bands simultaneously (230 and 690~GHz).  The first astronomical images
with one arcsecond angular resolution in the 690~GHz band have been
acquired.  Initial observations of sources with strong maser lines in
both bands demonstrate good correlation between the phases in the two
widely-separated frequency bands, at least on most of the antennas.
Although a number of instrumental issues with the SMA remain to be
explored and improved, this initial result does bode well for future
attempts at phase referencing the interferometer calibration from low
frequencies to high frequencies the submillimeter band (and perhaps
the terahertz band), and thereby improving the ultimate sensitivity of
traditional heterodyne interferometers at high frequencies.


\section*{Acknowledgment}
The authors would like to thank Irwin Shapiro for supporting the
development of the SMA from its inception.  We extend special 
thanks to those of Hawaiian ancestry on whose sacred mountain 
we are privileged to be guests.



%

\end{document}